\documentclass[fleqn,10pt]{SettingFiles/wlscirep}

\usepackage{float}
\usepackage{framed}
\usepackage{bm}
\usepackage{physics}
\usepackage{amsfonts}
\usepackage[utf8]{inputenc}
\usepackage[T1]{fontenc}
\usepackage{mathpazo} 
\usepackage{amsmath}
\usepackage{textalpha}
\usepackage{adjustbox}
\usepackage{makecell}
\usepackage{nicematrix}
\usepackage{array}
\usepackage{siunitx}

\usepackage{color, colortbl}
\newcommand{\changes}[1]{\textcolor{black}{#1}}

\definecolor{HeaderGray}{gray}{0.85}	
\definecolor{Gray}{gray}{0.9}

\newcommand{\nanomicrosec}{\SI{}{\nano\second}~$\sim$~\SI{}{\micro\second}}
\newcommand{\nanomilli}{\SI{}{\nano\second}~$\sim$~\SI{}{\milli\second}}
\newcommand{\microsec}{$\sim$~\SI{}{\micro\second}}
\newcommand{\nanosec}{$\sim$~\SI{}{\nano\second}}
\newcommand{\secs}{$\sim$~\SI{}{\second}}

\title{Perspective on unconventional computing using magnetic skyrmions}

\author[1]{Oscar Lee}
\author[2]{Robin Msiska}
\author[3]{Maarten A. Brems}
\author[3,a)*]{Mathias Kl\"aui}
\author[1,4,5,b)*]{Hidekazu Kurebayashi}
\author[2,c)*]{Karin Everschor-Sitte}

\affil[1]{London Centre for Nanotechnology, University College London, London, WC1H 0AH, United Kingdom}
\affil[2]{
Faculty of Physics and Center for Nanointegration Duisburg-Essen (CENIDE), University of Duisburg-Essen, 47057 Duisburg, Germany}
\affil[3]{Institut für Physik, Johannes Gutenberg-Universität Mainz, Staudingerweg 7, 55128 Mainz, Germany}

\affil[4]{Department of Electronic and Electrical Engineering, University College London, London, WC1E 7JE, United Kingdom}

\affil[5]{WPI Advanced Institute for Materials Research, Tohoku University, 2-1-1, Katahira, Sendai 980-8577, Japan}

\affil[*]{Authors to whom correspondence should be addressed: $^{\rm{a})}$~klaeui@uni-mainz.de, $^{\rm{b})}$~h.kurebayashi@ucl.ac.uk, $^{\rm{c})}$~karin.everschor-sitte@uni-duisburg-essen.de}

\begin{abstract}
Learning and pattern recognition inevitably requires memory of previous events, a feature that conventional CMOS hardware needs to artificially simulate. Dynamical systems naturally provide the memory, complexity, and nonlinearity needed for a plethora of different unconventional computing approaches. In this perspective article, we focus on the unconventional computing concept of reservoir computing and provide an overview of key physical reservoir works reported. We focus on the promising platform of magnetic structures and, in particular, skyrmions, which potentially allow for low-power applications. \changes{Moreover, we discuss skyrmion-based implementations of Brownian computing, which has recently been combined with reservoir computing. This computing paradigm leverages the thermal fluctuations present in many skyrmion systems. Finally,} we provide an outlook on the most important challenges in this field.
\end{abstract}

\begin{document}

\flushbottom
\maketitle

\section{Introduction to reservoir computing and magnetic skyrmions}

Modern-day applications of artificial intelligence (AI) have become pervasive in many aspects of our daily lives, and their importance is only predicted to increase. Artificial neural networks (ANNs), computational models inspired by the biological neural network architecture of the human brain, are primarily responsible for the rapid advancement of AI research~\cite{Goodfellow_2016, Schmidhuber_2015}. A class of ANN known as the recurrent neural network (RNN)~\cite{Sherstinsky_2020} excels at processing sequential or time series data. RNNs are distinguished by their "memory", which incorporates data from previous inputs to process a specific element of an input sequence.

\begin{figure*}[h]
\includegraphics[width=\linewidth]{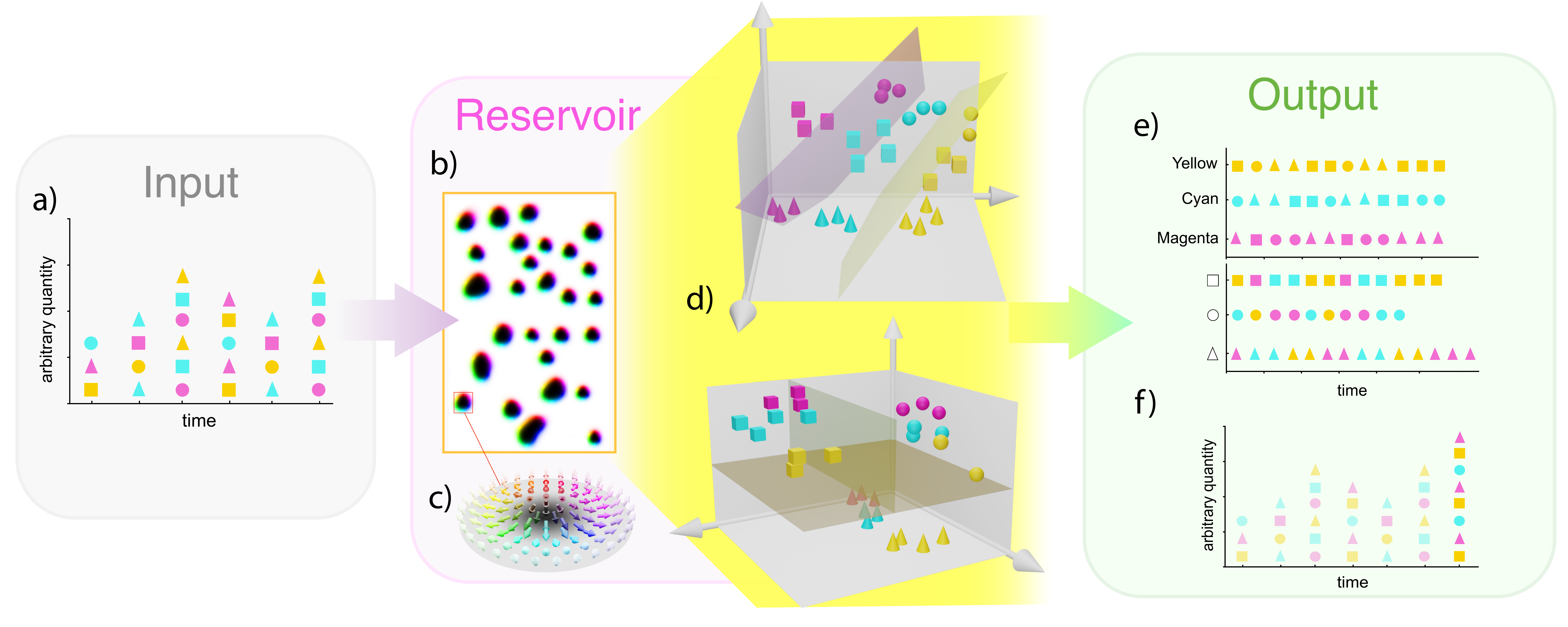}
\caption{\changes{Illustration of the reservoir computing framework taking the classification of a sequence of multicoloured assorted shapes as an example.} (a) An arbitrary temporal input signal \changes{(colourful shape sequence)} excites the (b) reservoir --- embodied in this illustration by a physical system made up of (c) magnetic skyrmions --- which then (d) projects the input data into a linearly separable higher dimensional space in which hyperplanes  can be used to (e) classify different desired features \changes{(shape or colour)} by only training the output readout. Moreover, the input-excited reservoir projection also enables other inference tasks, such as (f) time series prediction.}
\label{Fig:RC-summary}
\end{figure*}

Reservoir computing (RC) is a general and universal computational \changes{framework~\cite{Gonon_2020}} descended from RNNs. Its foundations can be traced back to two independently developed RNN-based models, echo-state networks by Jaeger~\cite{Jaeger_2001} and liquid-state machines by Maass et al.~\cite{Maass_2022}. \changes{RC consists of two main components: a fixed, randomly initialised nonlinear RNN system called the "reservoir" and a trainable readout layer. The reservoir, characterised by its recurrency and fading memory properties, acts as a high-dimensional, nonlinear projection of the input data, efficiently capturing the temporal information and inherent dynamics of the system. Recurrency in the reservoir enables it to maintain a continuous internal state, while the fading memory property ensures that more recent inputs have a higher impact on the reservoir states than older ones, allowing for efficient short-term memory. The higher dimensional mapping of inputs} enables spatiotemporal feature selection to be performed at the readout nodes using \changes{relatively simple methods such as regression algorithms (e.g. linear, ridge, and logistic regression)}. In contrast to conventional ANNs, which call for fine-tuning a plethora of interconnected node weights across multiple layers, the internal weights of the reservoir and the input nodes remain fixed, and only the weights of the readout nodes need to be trained. This significantly reduces the computational cost of learning, \changes{especially when compared to other RNNs}~\cite{Jaeger_2002_training}. Figure~\ref{Fig:RC-summary} schematically depicts a summary of the concept of reservoir computing. Here, the nonlinear reservoir is exemplified by a skyrmion fabric system.\cite{Prychynenko_2018,Bourianoff_2018,Pinna_2020, Msiska_2022}

For RC systems to work effectively, some crucial requirements need to be met: The reservoir needs to have high complexity. \changes{This complexity entails an intricate interplay of connectivity, dynamics, dimensionality, adaptivity, and nonlinearity. The reservoir's interconnected recurrent units exhibit rich and diverse interactions, while the system's dynamics capture complex input patterns. The number of its effective degrees of freedom must be larger than the dimensionality of the input. High dimensionality enhances learning and generalisation, while adaptivity allows the system to evolve during the learning process. Intrinsic nonlinearities enable the representation and processing of complex, nonlinear relationships in input data, making reservoir computing systems capable of handling sophisticated temporal and spatial data.} Another necessary characteristic is that the reservoir's internal state needs to be influenced by recent inputs while remaining unaffected by inputs from the distant past. This quality is referred to as fading/short-term memory~\cite{Jaeger_2002, Dambre_2012}. 
The extent of this fading memory has a profound effect on the information processing capacity of the reservoir~\cite{Carroll_2022, Dambre_2012}. Fading memory is what makes RC well-suited for processing temporal data with transient dependencies,~\cite{Maass_2022} such as stochastic or chaotic time series prediction~\cite{Jaeger_2004, Khovanov_2021}. Additionally, a reservoir needs to possess the ability to respond to a given input uniquely. More concretely, the reservoir should distinctly map the temporal history of a given input to a specific internal state. This is termed the echo state property~\cite{Jaeger_2001,Lukosevicius_2009,Yildiz_2012}. \changes{Formally, a reservoir is said to possess the echo state property if, for any input sequence $u(n)$ with time step $n$, the reservoir states $x(n)$ it generates satisfy the condition: For all $n > n_0$, where $n_0$ is the initial step, and for any pair of initial reservoir states $x(0)$ and $x'(0)$, the difference between the corresponding reservoir states $x(n)$ and $x'(n)$ vanishes as $n$ approaches infinity, i.e., $\lim_{n\to\infty} |x(n) - x'(n)| \rightarrow 0$.
}

Although RC does not train the internal weights of the reservoir, it is still often possible to optimise reservoir performance by tuning 
hyperparameters. For systems with many possible parameter choices, task-agnostic metrics help in identifying excellent hyperparameters~\cite{Ferreira_2009, Love_2021}.
Some reservoir systems exhibit dynamics that transition between non-chaotic and chaotic regimes upon adjustment of their intrinsic parameters.
Occasionally, such reservoirs can be optimised by adjusting them to operate at the so-called "edge of chaos" --- a critical phase transition point beyond which the reservoir system's dynamics become chaotic~\cite{Langton_1990,Bertschinger_2004}. While this approach can be effective in designing reservoirs with chaotic tendencies, it is not universally applicable, and there have been exceptions to this hypothesis~\cite{Legenstein_2007, Carroll_2020, Jaeger_2023}. For this reason, it is imperative to understand the dynamical trends of a chosen reservoir system.

A branch of RC called \textit{physical} RC (PRC) has emerged, in which physical systems are used as reservoirs~\cite{Tanaka_2019,Nakajima_2020,Nakajima_2021_book,Christensen_2022}. Physical systems often naturally fulfil the RC criteria of being complex, nonlinear, and possessing a short-term memory. \changes{It is crucial to note that for physical reservoirs, consistent reproducibility is an additional essential prerequisite, ensuring outcomes can be replicated under similar conditions. Reproducibility also entails that the reservoir is robust against noisy fluctuations and other internal transient dynamics that don't promote the nonlinear transformation of inputs but persist even after input signals are removed.} A PRC measure of fading memory needs to include these fleeting dynamics as they have been shown to support short-term memory~\cite{Buonomano_2009,Dambre_2012,Nokkala_2022}.

In this Perspective, we describe reservoirs epitomised by skyrmions such as the skyrmion fabric system shown in Fig.~\ref{Fig:RC-summary}.  Magnetic skyrmions are localised magnetic whirls possessing particle-like properties. Skyrmions were proposed by Tony Skyrme in 1961 in a model describing elementary particles~\cite{Skyrme_1961}. The magnetic versions studied nowadays were theoretically predicted in 1989~\cite{Bogdanov_1989} and experimentally observed in 2009~\cite{Muehlbauer_2009}. Since then, they have been shown to occur in many magnetic systems ranging from insulators to metals at various temperatures, even above room temperature~\cite{Fert_2017,Everschor-sitte_2018,Bogdanov_2020,Gobel_2021,Jiang_2017,Finocchio_2016,Tokura_2021}. 
Furthermore, skyrmions have been observed
as singular objects~\cite{Yu_2010,Woo_2016,Moreau-Luchaire_2016,Boulle_2016}, clusters~\cite{Leonov_2014,Du_2015}, skyrmion lattices~\cite{Muehlbauer_2009,Adams_2011,Heinze_2011}and in the form of intermediate skyrmion phases known as skyrmion fabrics (Fig.~\ref{Fig:RC-summary}b)~\cite{You_2015}.
An example of a single skyrmion is illustrated in Fig.~\ref{Fig:RC-summary}c.

Skyrmions have a non-trivial topology which implies that a topological invariant can be associated with them, quantified by the topological index
\begin{equation}
Q = \frac{1}{4\pi}\int\textbf{M}\cdot\left(\frac{\partial\textbf{M}}{\partial x}\times \frac{\partial\textbf{M}}{\partial y}\right)\mathrm{d}x\mathrm{d}y,
\end{equation}
where \textbf{M} is the magnetisation unit vector, and the integral is taken over a two-dimensional space. $Q=\pm 1$ for skyrmions in particular. Topology has a profound effect on the physics of skyrmions and influences phenomena such as transport~\cite{Schulz_2012} and thermal~\cite{Schuette_2014, Zazvorka_2019, Zhao_2020, Nozaki_2019} motion. Topology massively increases the skyrmion's robustness against structural defects and impurities~\cite{Mueller_2015}. Topological properties can be exploited to build stable reservoir systems~\cite{Prychynenko_2018,Bourianoff_2018}. 

Magnetic spin textures such as skyrmions originate from the interplay of different competing energetic contributions within a given system~\cite{Bogdanov_2020}. Typically, there is at least one energy term that favours a uniform ferromagnetic configuration of magnetic moments, e.g. exchange or anisotropy, and there are other terms that promote twisted configurations, e.g. Dzyaloshinskii–Moriya interaction (DMI) and demagnetisation~\cite{Everschor-sitte_2018}. In magnetic materials with broken inversion symmetry, such as uniaxial non-centrosymmetric ferromagnets~\cite{Rossler_2006}, cubic helimagnets~\cite{Wilson_2013} and thin film heterostructure systems with structural inversion asymmetry~\cite{Everschor-sitte_2018}, DMI effects~\cite{Magni_2022} become pronounced. \changes{The aforementioned thermally activated dynamics also enable skyrmion-based Brownian computing approaches~\cite{Raab_2022,Peper_2013,Lee_2016,Jibiki_2020,Brems_2021,Zazvorka_2019}, as discussed in section 3.}

Skyrmion dynamics can be manipulated by spin torques~\cite{Jonietz_2010, Zang_2011, Schulz_2012, Iwasaki_2013}, magnetic fields~\cite{Zhang_2018, Moutafis_2009,Buttner_2015}, electric fields~\cite{Nakatani_2016,Liu_2019,Ba_2021}, magnons~\cite{Zhang_2015, Zhang_2017,Psaroudaki_2018}, temperature gradients~\cite{Kong_2013,Raimondo_2022} and thermal fluctuations~\cite{Mochizuki_2014, Troncoso_2014,Wang_2020}. Through such mechanisms, it is possible to reliably control the creation (writing)~\cite{Romming_2013}, detection (reading)~\cite{Romming_2013}, rotation~\cite{Everschor_2011, Everschor_2012} and even annihilation (deleting)~\cite{Romming_2013} of magnetic skyrmions. These qualities, along with topological stability, small size~\cite{Wang_2018,Wu_2021}, manoeuvrability around material defects~\cite{Lin_2014} and ultra-low power operation~\cite{Jonietz_2010,Schulz_2012} pave the way for magnetic skyrmions to be used in applications such as skyrmion-based racetrack memories~\cite{Fert_2013,Tomasello_2014}, logic devices~\cite{Zhang_2015_logic,Luo_2018}, magnetic tunnel junctions~\cite{Zhang_2018_MTJ,Penthorn_2019}, nano-oscillators~\cite{Zhang_2015_osc,Garcia-sanchez_2016}, and unconventional computing schemes~\cite{Grollier_2020,Zhou_2021,Finocchio_2021,Brems_2021} like neuromorphic~\cite{Song_2020}, probabilistic~\cite{Pinna_2018,Zazvorka_2019}, and reservoir computing~\cite{Prychynenko_2018,Raab_2022,Lee_2022}.

\section{Overview of key physical reservoir works}

Systems from diverse disciplines have demonstrated their capabilities of constructing a physical reservoir~\cite{Tanaka_2019,Nakajima_2020,Nakajima_2021_book,Christensen_2022}, including bioelectronics~\cite{Cucci_2021,Sumi_2022,Rashid_2021,Pecqueur_2018}, electronics~\cite{Toprasertpong_2022,Elbedwehy_2022,Nowshin_2020,Joksas_2022,Christensen_2022}, magnonics~\cite{Watt_2020,Millet_2021,Lee_2022,Korber_2022}, memristors~\cite{Milano_2021,Zhong_2022,Mehonic_2020,Kumar_2022}, \changes{nanomagnets~\cite{Nomura_2019,Dawidek_2021,Welbourne_2021,Gartside_2022,Stenning_2022,Vidamour_2022}}, photonics~\cite{derSande_2017,Antonik_2017,Takano_2018,Brunner_2019,Rafayelyan_2020,Xiang_2021,Nakajima_2021}, and spintronics~\cite{Torrejon_2017,Romera_2018,Grollier_2020,Zhou_2021,Finocchio_2021,Joksas_2022,Raab_2022,Namiki_2022,Yokouchi_2022}, summarised in Table~\ref{tab:Table1}. This yields an advantage from an application standpoint as different reservoirs could be designed pertinent to a specific application considering their available inputs and readout mechanisms~\cite{Cucchi_2022}. 

\begin{table}[t!]
\centering
\captionsetup{justification=raggedright,singlelinecheck=false}
\caption{Examples of experimental and general skyrmion reservoir systems. Note that the 'Timescale' column indicates the system’s intrinsic timescales, while the operation speeds may be limited by their measurement and control schemes. See main text for abbreviations.}
\normalsize
\label{tab:Table1}
\begin{adjustbox}{width=\textwidth}
\begin{NiceTabular}{
  >{\raggedright\arraybackslash}m{1.0cm}
  >{\raggedright\arraybackslash}m{2.0cm}
  >{\centering\arraybackslash}m{2.0cm}
  >{\centering\arraybackslash}m{2.3cm}
  >{\centering\arraybackslash}m{2.3cm}
  >{\centering\arraybackslash}m{2.3cm}
  >{\centering\arraybackslash}m{1.4cm}
  >{\centering\arraybackslash}m{3.0cm}
}

\hline
\rowcolor{HeaderGray}
\rule{0pt}{12pt}\raisebox{0.25ex}{\textbf{Ref.}} & \rule{0pt}{12pt}\raisebox{0.25ex}{\textbf{Discipline}} & \rule{0pt}{12pt}\raisebox{0.25ex}{\textbf{Reservoir}} & \rule{0pt}{12pt}\raisebox{0.25ex}{\textbf{Input}} & \rule{0pt}{12pt}\raisebox{0.25ex}{\textbf{Output}} & \rule{0pt}{12pt}\raisebox{0.25ex}{\textbf{Readout}} & \rule{0pt}{12pt}\raisebox{0.25ex}{\textbf{Timescales}} & \rule{0pt}{12pt}\raisebox{0.25ex}{\textbf{Demonstration}}\\
\hline
\hline
\multicolumn{8}{l}{\rule{0pt}{12pt}\raisebox{0.25ex}{\textbf{Experimental physical reservoir systems}}}\\
\hline
\hline

\rowcolor{Gray}
[\citenum{Sumi_2022}] & Bioelectronics & mBNN & Photo-stimulations & Evoked activity & Fluorescent calcium imaging & \secs & Spoken digit recognition\\

[\citenum{Cucci_2021}] & Bioelectronics & OECT & Voltage waveforms & Voltage & Analog DAQ system & \secs & Heartbeat classification\\  

\rowcolor{Gray}
[\citenum{Toprasertpong_2022}] & Electronics & FeFET & Voltage waveforms & Current & Terminal currents & \nanomicrosec & Temporal-XOR \& parity-check\\

[\citenum{Zhong_2022}] & Memristors & DM & Voltage pulses & Current & RRAM & \microsec & Arrhythmia detection \& gesture recognition\\

\rowcolor{Gray}
[\citenum{Milano_2021}] & Memristors & NW networks & Voltage pulses & Voltage & RRAM & \microsec & Handwritten digit classification\\

[\citenum{Vidamour_2022}] & Nanomagnets & Nanorings & Rotating magnetic fields & AMR response & Electrical contacts & \nanomicrosec & Signal transformation \& Spoken digit recognition\\

\rowcolor{Gray}
[\citenum{Gartside_2022,Stenning_2022}] & Nanomagnets & ASVI & Magnetic field waveforms & Spinwave spectra & FMR & \nanosec & Signal transformation \& forecasting\\

[\citenum{Torrejon_2017}] & Spintronics & STNO & Voltage waveforms & Voltage & Diode rectification & \microsec & Spoken-digit recognition\\

\hline
\hline
\multicolumn{8}{l}{\rule{0pt}{12pt}\raisebox{0.25ex}{\textbf{Skyrmion-based physical reservoir systems (*experimental works)}}}\\
\hline
\hline

\rowcolor{Gray}
[\citenum{Lee_2022}] & Magnonics* & Chiral magnet & Magnetic field waveforms & Spinwave spectra & FMR & \nanosec & Signal transformation \& forecasting\\

[\citenum{Raab_2022}] & Spintronics* & Confined skyrmions & Voltage pulses & Skyrmion displacement & Kerr-microscopy / electrical contacts & \nanomilli & Boolean logic operations\\

\rowcolor{Gray}
[\citenum{Yokouchi_2022}] & Spintronics* & Hall bars & Magnetic fields & Anomalous Hall voltage & Electrical contacts &  \nanomicrosec & Waveform \& handwritten digit recognition\\

[\citenum{Prychynenko_2018,Bourianoff_2018,Pinna_2020,Msiska_2022}] & Spintronics & Skyrmion fabrics & Voltage waveforms & AMR response, local magnetisation  & Spatially resolved magnetisation & \nanosec & Temporal pattern recognition, spoken digit recognition\\

\rowcolor{Gray}
[\citenum{LeeMu_2022}] & Spintronics & Thin plate & Microwave pulses & Magnetisation oscillations & Oscillation detectors & \nanomicrosec & Short-term memory \& parity-check\\

[\citenum{Jiang_2019}] & Spintronics & MSM & Current pulses & Skyrmion position & Mathematical function & \nanosec & Handwritten digit classification\\

\rowcolor{Gray}
[\citenum{Rajib_2022}] & Straintronics & Thin film & Voltage-induced
strains & Time-resolved magnetisation & MTJ & \nanosec & Short-term memory \& parity-check\\

\hline

\end{NiceTabular}
\end{adjustbox}
\end{table}

In bioelectronic systems, information is processed using biocompatible materials or organic biological architectures in conjunction with electronic sensors. While many approaches primarily focus on clinical applications, their use for neuromorphic computation has begun gaining interest~\cite{Lee_2021,Song_2022,Kim_2022}. A study by Sumi et al.~\cite{Sumi_2022} has demonstrated RC on a micropatterned biological neuronal network (mBNN). Here, the input data were transformed to frequency-dependent photostimulation to create the reservoir using optogenetic techniques supplied to rat cortical neurons grown on micropatterned substrates. The readout mechanism incorporated measurements by fluorescent imaging via calcium probes \changes{(20 frames/second)}, where the spontaneous and evoked activities of mBNN were trained \changes{to} demonstrate spatial pattern and spoken digit recognition tasks. \changes{The performance of mBNN reservoirs are typically bound to its timescales of short-term memory capacity, which ranges from tens of milli to a few seconds~\cite{Dranias_2013,Kubota_2019,Sumi_2022}. While it may not be suitable for high-speed electronic applications, it may apply in specific cases where biological timescales share similar orders~\cite{Venezia_2015,Sumi_2022}}. On the other hand, organic electrochemical transistors (OECTs) have also sparked promises for RC~\cite{Rashid_2021,Cucchi_2022}. Using data-encoded voltage waveforms as inputs and analogue readings of output time-variant voltages, examples of \changes{nonlinear} signal and heartbeat classification have been shown~\cite{Pecqueur_2018,Cucci_2021}.

Diverse electronic systems, including analogue circuits, FPGAs, memristors and ferroelectrics, highlight their potential for RC~\cite{Nowshin_2020,Joksas_2022,Christensen_2022}. Such systems have flexible scalability and pose a benefit of circuit-level implementation, compatible with silicon-based CMOS technologies at low power. In particular, memristive technologies have continued to proliferate for their role in neuromorphic computation and RC~\cite{Mehonic_2020,Kumar_2022}. Diverse architectural designs have been proposed, and detailed studies have investigated fundamental properties in building or improving the system efficiency/performance of the reservoirs. Numerous experiments have explored various design paradigms that take a step closer to device-level implementations of RC. For example, a recent work by Milano et al.~\cite{Milano_2021} uses self-organised nanowire (NW) networks with a memristive architecture as a reservoir. After inputting a sequence of voltage pulses that encodes the data, it utilises resistive random access memory (RRAM) as a readout mechanism to convert the output voltages from the NW networks to a matrix of currents that could be trained. In this study, handwritten digit recognition and signal forecasting have been demonstrated. Similarly, a fully-analogue RC~\cite{Zhong_2022} involving non-volatile memristor arrays used as RRAM in the readout layer with data-translated input voltage pulses to dynamical memristors (DM) was shown to allow the detection of arrhythmia and dynamic hand gesture recognition. The study reports that the power consumption of such a system comprising 24 DMs is 22.2~$\mu$W. 

Among silicon-compatible systems, ferroelectric field-effect transistors (FeFET) have been realised as an alternative approach in designing future electronic components for in-memory~\cite{Khan_2020} and neuromorphic computation~\cite{Markovic_2020,Kim_2021,Mulaosmanovic_2021}. As a multi-terminal device, while FeFETs share similar nature with standard field-effect transistors, it uses ferroelectric materials for the gate insulator. Its nonlinearity stems from the time-dependent polarisation reversal process on the input gate voltage. Harnessing this property allows the output currents to exhibit history-dependent and nonlinear dynamics adequate for RC~\cite{Toprasertpong_2022}. On this note, by measuring the drain, source and substrate currents from a data-mapped voltage waveform input, Toprasertpong et al.~\cite{Toprasertpong_2022} have performed RC on a HfO$_2$-based FeFET to compute logic-based tasks, including temporal-XOR and parity-checks. \changes{While FeFETs are one example, semiconductor electronics provide additional room for increasing the fabrication complexity with innovative engineering solutions and may lead to large-scale integration~\cite{Salahuddin_2018}.}

Spintronic \changes{and magnonic} systems have also highlighted advantages as a physical platform for RC\changes{~\cite{Grollier_2016,Grollier_2020,Finocchio_2021,Allwood_2023}}. Works by Torrejon et al.~\cite{Torrejon_2017} demonstrated PRC using a network of spin-torque nano oscillators (STNOs) made from magnetic tunnel junctions (MTJs) for spoken digit (pattern) recognition, with an accuracy of up to 99.6~\% and nonlinear waveform classification tasks. The data-mapped voltages were input as currents into the STNOs, and the output response of rectified time-dependent voltages was recorded to construct a reservoir. \changes{Subsequently, the work has sparked interest in further developments of PRC with STNOs and MTJs~\cite{Furuta_2018,Romera_2018,Kanao_2019,Markovic_2019,Jiang_2019,Tsunegi_2019,Yamaguchi_2023}. Diverse nanomagnetic systems, including the use of their magnetic dipole-coupling interactions, surface acoustic waves, artificial spin-vortex ice (ASVI), and nanoring arrays, have also proposed and shown promising PRC performances by exploiting their high-dimensionalities and rich nonlinear dynamics~\cite{Nomura_2019,Dawidek_2021,Welbourne_2021,Gartside_2022,Chowdhury_2022,Stenning_2022,Vidamour_2022}. For example,} Gartside et al.~\cite{Gartside_2022} utilised nanomagnetic arrays by applying magnetic field inputs and measuring the nonlinear nucleation dynamics of spinwaves spectra \changes{using ferromagnetic resonance (FMR)}. \changes{Dawidek et al.~\cite{Dawidek_2021} proposed PRC by manipulating the domain wall (DW) population in the nanoring arrays by rotating the applied magnetic fields. It was later experimentally demonstrated by Vidamour et al.~\cite{Vidamour_2022} through transport measurements collecting anisotropic magnetoresistance (AMR) signals associated with the annihilation and repopulation of the DWs in the nanorings.} Furthermore, complex magnetic structures inherently provide all the important components of a reservoir without creating a system of interconnected neurons: complexity, nonlinearity, and short-term memory. 



\changes{On the other hand, various skyrmion-based PRC systems have been proposed~\cite{Prychynenko_2018,Bourianoff_2018,Jiang_2019,Pinna_2020,Msiska_2022,Rajib_2022,LeeMu_2022} and experimentally demonstrated~\cite{Yokouchi_2022,Raab_2022,Lee_2022}. These systems  may offer particular advantages, including speed, energy efficiency, task adaptability, and scalability, as high-speed ($\sim$~nanosecond timescales) and low-power ($\sim$~microwatt or less) alternatives to existing PRC schemes~\cite{Prychynenko_2018,Bourianoff_2018,Pinna_2020,Raab_2022}. In particular, rich controllability of material parameters can lead to performance improvements by adjusting the system's reservoir properties and reconfiguring its nonlinearity and memory capacity~\cite{Bourianoff_2018,Pinna_2020,Love_2021,Lee_2022}. For example, an experimental demonstration by Lee et al.~\cite{Lee_2022} has shown that adding skyrmions to conical/ferromagnetic magnetic phase-reservoirs can enhance memory in the system and improve forecasting tasks by an order of magnitude. Raab et al.~\cite{Raab_2022} highlighted the potential for device down-scaling by exploiting current-induced spin-orbit torques to manipulate skyrmions and using TMR to detect the presence of skyrmions in skyrmion-based RCs. Yokouchi et al.~\cite{Yokouchi_2022} utilised magnetic field-induced skyrmion dynamics in Hall bar arrangements and showed promising handwritten digit recognition tasks with anomalous Hall voltage measurements.} 

\changes{Further theoretically proposed RC schemes predict promising results. Spintronic RCs include the voltage-dependent skyrmion positions in a magnetic skyrmion memristor (MSM)~\cite{Jiang_2019}, exploiting the resistance or magnetisation changes in skyrmion fabrics~\cite{Prychynenko_2018,Bourianoff_2018,Pinna_2020,Msiska_2022} and measuring the spinwave propagations in a thin plate magnet hosting skyrmions~\cite{LeeMu_2022}. A straintronic skyrmion-based PRC system proposes utilising nonlinear breathing dynamics of skyrmions via voltage-induced strain in an MTJ block~\cite{Rajib_2022}. Therefore, given appropriate material choices and advancements in device engineering, such properties of magnetic skyrmions could be manoeuvred for competitive performance in unconventional computing schemes. Nevertheless, addressing fabrication complexities, controlling the interplay between pinning effects, thermal fluctuations, and skyrmion dynamics, and improvements for faster readout mechanisms (currently, most demonstrations are confined to the limits of measuring instruments) are crucial for realising its full potential~\cite{Bourianoff_2018,Pinna_2020,Raab_2022,Lee_2022}. More in-depth examples of some of the above skyrmion-based unconventional systems are discussed in the following section.}

\section{Skyrmion-based reservoir and Brownian computing}

Skyrmion-based RC, working at the nanosecond timescale with power consumption in the microwatt regime, has been theoretically proposed by Prychyneko et al.~\cite{Prychynenko_2018}. Due to its independence of concrete details of the reservoir, the input and readout method, various approaches and models have been predicted and analysed~\cite{Pinna_2020,Bourianoff_2018,Jiang_2019,Lee_2022_spin,Rajib_2022, Msiska_2022}.
For example, Prychynenko et al.~\cite{Prychynenko_2018}, Bourianoff et al.~\cite{Bourianoff_2018}, Pinna et al.~\cite{Pinna_2020}, Raab et al.~\cite{Raab_2022} and Msiska et al.~\cite{Msiska_2022} have studied the response of a skyrmion fabrics system (as exemplified in Fig.~\ref{Fig:RC-summary}b) to voltage inputs. In these studies, reservoir computing is based on exploiting the nonlinear current-voltage characteristics of skyrmion systems due to the complex interplay of current-induced dynamics and pinning effects. A readout is possible, for example, as a time-traced resistance signal, a spatially resolved magnetisation measurement, or a combination of both. 
By adjusting material parameters that appear as hyperparameters in the reservoir computing model, the skyrmion reservoir can be customised for tasks that rely more on memory or nonlinearity~\cite{Love_2021}.
For example, the recently simulated multi-dimensional input skyrmion-based reservoir \changes{demonstrated best-in-class in-materio RC performance in a standard spoken digit classification benchmark task}~\cite{Msiska_2022}. 
Raab et al. combined the RC principle with the Brownian computing (BC) concept and demonstrated a skyrmion-based RC experimentally~\cite{Raab_2022}. BC refers to the broad idea of exploiting intrinsic random dynamics of a physical system to the benefit of a computing architecture~\cite{Raab_2022,Peper_2013,Lee_2016,Jibiki_2020,Brems_2021}. It is inspired by noise-exploiting mechanisms in biological processes where, e.g., Brownian motion drives molecular machines~\cite{Yanagida_2008,Reimann_2002} – hence the name. \changes{There are two main conditions for the underlying system to transfer the advantages of these biophysical mechanisms to computing devices: First, the system must exhibit significant thermal dynamics at operating temperature, which is typically room temperature. Second, for good integrability in existing computing hardware, the system must be addressable electrically, i.e., in- and outputs may be set and read-out by electrical means.} Skyrmions are a particularly promising system for BC as they have been shown to undergo thermally activated diffusion~\cite{Zazvorka_2019,Kerber_2021,Nozaki_2019} \changes{and can be measured and manipulated by a variety of different mechanisms.\cite{Jonietz_2010,Zang_2011, Schulz_2012,Iwasaki_2013,Zhang_2018, Moutafis_2009,Buttner_2015,Nakatani_2016,Liu_2019,Ba_2021,Zhang_2015, Zhang_2017,Psaroudaki_2018,Kong_2013,Raimondo_2022,Romming_2013}}. In addition, the thermal effects compete with different skyrmion interactions \changes{and drives} at room temperature\changes{~\cite{Gruber_2022,Ge_2022,Raab_2022}}. There exists a multitude of ways how exactly thermal random effects can be exploited for different computing architectures~\cite{Raab_2022,Zazvorka_2019,Jibiki_2020,Brems_2021,Ishikawa_2021}. In particular, RC can be realised by combining the RC concept with thermally activated skyrmion dynamics~\cite{Raab_2022}, which is discussed in the next subsection, along with other skyrmion-based RC approaches. The two subsequent sub-sections then introduce other non-conventional computing approaches based on thermally activated skyrmion dynamics. These approaches, apart from being promising future applications on their own, aid in understanding the versatility of the BC concept in combination with different computing architectures such as RC. 

\subsection{Reservoir computing using skyrmions}

\begin{figure*}[h]
    
\includegraphics[width=\linewidth]{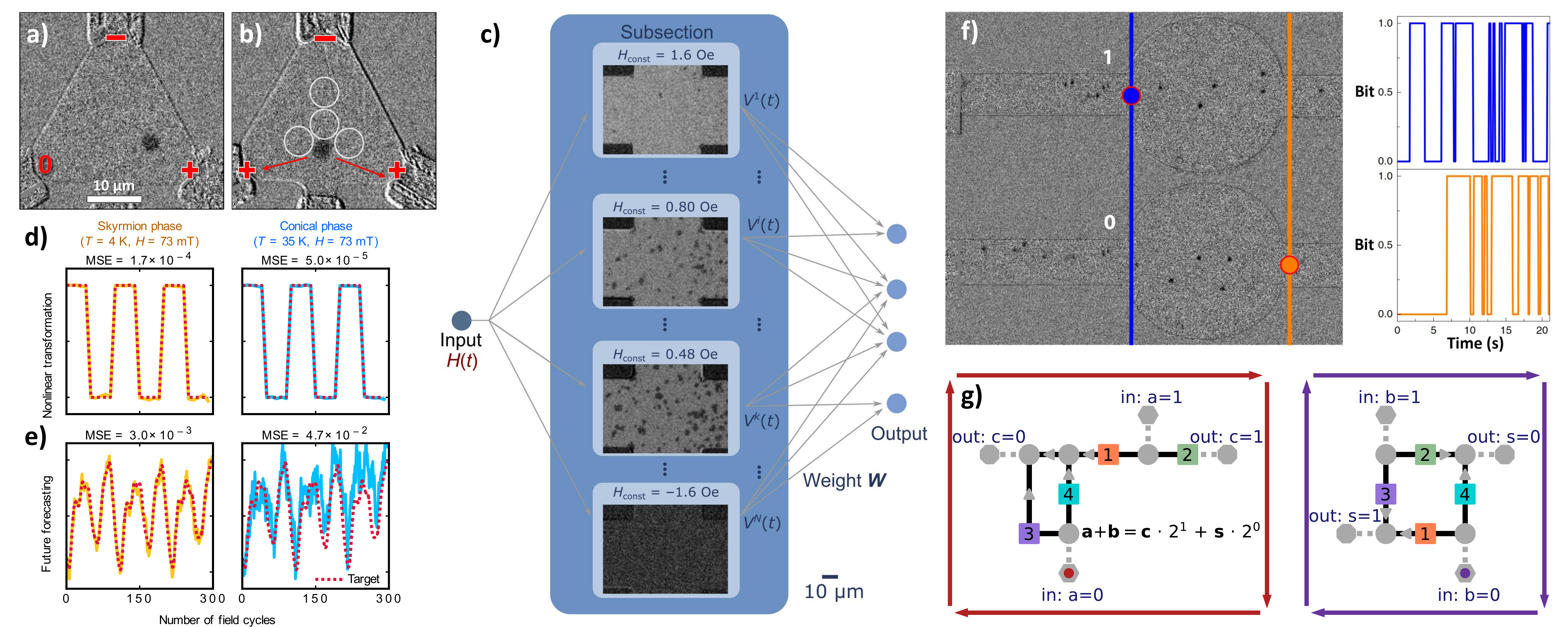}

    \caption{a)\& b) Kerr-microscopy images of the skyrmion-based Brownian RC device by Raab et al.~\cite{Raab_2022} for different input voltage combinations. The white circles in b) depict the regions within which the local skyrmion occupation probability is measured. Adapted from Ref.~\cite{Raab_2022} (CC BY 4.0). c) Schematic illustration of the skyrmion-based RC by Yokouchi et al.~\cite{Yokouchi_2022}, including Kerr-microscopy images of the Hall bars containing skyrmions at different constant OOP fields $H_{const}$. Adapted from Ref.~\citenum{Yokouchi_2022} (CC BY-NC 4.0). d)\& e) Performance comparison of RC for a d) nonlinear transformation and e) future forecasting tasks using skyrmion (orange) and conical (blue) magnetic phase spaces, respectively. Different magnetic modes are accessed by changing the applied temperature to the magnet. f) Kerr-microscopy image of the skyrmion reshuffler device by Zázvorka et al.~\cite{Zazvorka_2019} The input signal is constructed as a time frame in which the skyrmion crosses the blue threshold line. The output is produced on crossing the orange line. The corresponding input signal is depicted in blue (top), and the resulting output signal in orange (bottom). Adapted from Ref.~\citenum{Zazvorka_2019} with permission from the authors. Copyright Zázvorka et al. under exclusive license to Springer Nature Limited 2019. g) Circuit layout for a Brownian token-based half-adder by Brems et al.~\cite{Brems_2021}. Cjoins (coloured/numbered squares) can only be passed by both signal carriers (\changes{red and violet} bold dots) together. \changes{The current input placement of the tokens indicates an input a+b=0+0. The red token on the left side may take computational paths which lead either to Cjoin 3 or 4, whereas the violet token on the right side may reach Cjoins 1 and 4. Since Cjoins can only be traversed in pairs of two tokens, only Cjoin 4 can be passed. By similar tracing of the accessible and computational forward paths, it becomes clear that a value of 0 in both output digits is the only possible result, even if the token movement is completely random.} The coloured arrows show the relevant directions of driven motion to tune the balance of computation speed and energy consumption of the device. Adapted with permission from Brems et al. Ref.~\citenum{Brems_2021}. Copyright 2021 AIP Publishing LLC.}
    \label{fig:Figure3}

\end{figure*}

While several of the theoretical skyrmion RC concepts~\cite{Prychynenko_2018,Pinna_2020,Bourianoff_2018} heavily rely on the presence of local pinning sites~\cite{Gruber_2022}, Raab et al. realised Brownian RC by overcoming pinning effects in confined geometries by using thermal skyrmion diffusion~\cite{Raab_2022}. Their proof-of-concept reservoir consists of a single skyrmion in a triangular confinement~\cite{Song_2021}, which provides an automatic reset mechanism after the operation due to the repulsive skyrmion-boundary interaction~\cite{Ge_2022}. Inputs are encoded by the patterns of voltages at the corners of the triangle (Fig.~\ref{fig:Figure3}a\&b). The resulting current distribution acts as a biasing-mechanism for the thermal skyrmion motion and thereby alters the average spatial distribution of the skyrmion. In a nano-scale device, one can exploit that the probability for a certain region of the sample to be occupied by a skyrmion determines the average local tunnel magnetoresistance (TMR)~\cite{Tomasello_2017}. For the proof-of-concept device, the local occupation probability was determined in four regions (white circles in Fig.~\ref{fig:Figure3}b) using Kerr-microscopy to mimic TMR readout via magnetic tunnel junctions~\cite{Tomasello_2017}. The study demonstrates that training a linear readout based on the local occupation probabilities already suffices this minimalistic Brownian RC device to perform 2- and 3-input logic operations including the nonlinearly separable XOR. Moreover, exploiting thermal effects allows for ultra-low-current operation and overcomes pinning effects which would hinder proper operation in a diffusion-free system. \changes{Distinguishability of the systems’ responses to different input stimuli is key for reliable operation. Pinning effects can drastically reduce the output configuration space as the skyrmion can become pinned at the same position for various excitations hindering proper operation, even if the pinning strength is only slightly stronger than the strength of the drive. This effect can be mitigated by employing thermally active skyrmions, as the resulting skyrmion distribution still reflects both pinning and drive, given that both energy scales are comparable to the scale of thermal fluctuation.}

Another skyrmion-based RC concept was realised by Yokouchi et al., that have experimentally studied a skyrmion-based RC device capable of complex pattern recognition~\cite{Yokouchi_2022}. Their reservoir consists of a collection of Hall bars containing skyrmions, each at a different constant out-of-plane (OOP) field (Fig.~\ref{fig:Figure3}c). The input signal is encoded as a time-dependent OOP field in addition to the constant OOP field. Due to the different constant OOP fields, the skyrmions in each Hall bar potentially react differently to the input signal. Training is then performed to tune the weights combining the anomalous Hall voltages of all Hall bars. The reservoir succeeds in high-accuracy handwritten digit recognition as well as waveform recognition.

The spectral properties of skyrmions can be exploited for constructing a physical reservoir. Lee et al. recently demonstrated that the GHz dynamics of skyrmions generated in \changes{the class of chiral magnets can provide a scheme of phase-tunable, task-adaptive PRC~\cite{Lee_2022}.} By utilising rich thermodynamical phases available in \changes{multiferroic} Cu$_2$OSeO$_3$ \changes{at low temperatures}, they show that the single reservoir unit can offer multiple reservoir properties. Hence, adaptive to different computational tasks that require different reservoir \changes{metrics} (i.e., nonlinearity, memory capacity \& complexity). \changes{The studied scheme is shown to be transferable to other similar systems, including Co$_{8.5}$Zn$_{8.5}$Mn$_{3}$ and FeGe, to operate at above and near room temperatures}. 
In their work, external magnetic field values and temperature are controlled to navigate between available phase spaces to modify the key reservoir properties on demand. Subsequently, translating the input data (e.g. a sinewave signal) into a sequence of magnetic field values can encode its information to the spectral states of various spinwave modes to construct a reservoir. During this process, a particular input protocol, named ``mapped field-cycling", is incorporated to nucleate metastable magnetic phase spaces such as low-temperature skyrmions~\cite{Halder_2018,Aqeel_2021,Lee_2021_GHz}. The study demonstrates that the computational power for different tasks highly depends on the choice of the magnetic phase space. For example, as shown in Fig.~\ref{fig:Figure3}d\&e, while the skyrmion textures excel in future forecasting, their performance deteriorates substantially for linear-to-nonlinear transformation tasks. However, the conical modes observe the opposite behaviour, suggesting a correlation between the intrinsic magnetic phase properties and the task capability, which is related to the properties of the reservoir. These results highlight that a single material system can be reconfigured (by adjusting the magnetic field or temperature) based on the task's nature without creating alternative reservoir systems each time and takes a step closer to flexible on-demand PRC. The task-adaptive nature is important since a typical physical reservoir is often fixed and inflexible in terms of reservoir properties due to the constraint of specific response phenomena of a given physical system: consequently, many physical reservoirs result in severely constrained computational performance as it lacks the versatility to meet demands requiring different reservoir properties. For example, a reservoir system constructed with a high nonlinearity would not be adequate in performing tasks requiring a high memory capacity, and vice versa.

\subsection{Stochastic computing using skyrmions}

 Thermal excitations can not only be used in reservoir computing~\cite{Raab_2022} but also for stochastic computing. In skyrmion-based stochastic computing~\cite{Pinna_2018,Zazvorka_2019,Zhang_2020}, numerical values are encoded as the probability of a “1” occurring in random bit-streams and computation results are obtained by averaging over the results of bit-wise operations. For instance, multiplication can be realised by bit-wise application of the AND operation on two input-streams since the “1”-probability in the resulting bit-stream is equal to the product of the “1”-probabilities of the two input-streams. Skyrmion-based stochastic computing can offer interesting advantages such as high error tolerance and increasing accuracy  with computation time (length of processed bit-stream). However, one key challenge in this approach is that strong correlations between bit streams severely impair computations as the individual bit-wise computation results will no longer inherit the correct statistical distribution from the input-streams. In 2019, Zázvorka et al. experimentally constructed a skyrmion-based Brownian reshuffler device to decorrelate bit-streams~\cite{Zazvorka_2019}. As shown in Fig.~\ref{fig:Figure3}f, the device consists of an upper and a lower channel, and a skyrmion passing a certain section of a channel at a given time sets the bit-stream’s value to “1” or “0”, respectively. Each channel contains a chamber where the effects of the overall current-induced drift to the right are combined with thermal random diffusion to decorrelate the pre-chamber (blue) and post-chamber (orange) bit-streams. It was demonstrated that this reshuffler device leads to very good decorrelation of the bit-streams while keeping the value encoded in the bit-stream constant.\\
 
\subsection{Brownian token-based computing using Skyrmions}

In Brownian token-based computing~\cite{Peper_2013, Lee_2016,Jibiki_2020,Brems_2021, Nozaki_2019}, computations are performed \changes{as discrete and indivisible signal carriers (tokens) explore a network of computational paths by means of random motion. Therein, all logic is contained within the circuit layout such that random motion suffices for the tokens to find the computational forward paths to the elements, which advance the computation. Magnetic skyrmions are particularly promising token-candidates due to their quasi-particle nature and thermally induced diffusive dynamics, which can be easily manipulated.} The key advantage of this computing method is that energy must not be invested to move the information carriers but only to synchronise their movement at certain points in the circuit, potentially allowing for low-energy operation. Note, matching coloured/labelled squares in Fig.~\ref{fig:Figure3}g represent Cjoin modules which can only be passed across by two tokens at the same time. Brems et al. has proposed a skyrmion-suitable crossing-free circuit for a Brownian half-adder (Fig.~\ref{fig:Figure3}g) along with a framework to tune the balance of speed and energy consumption of token-based computers using artificial diffusion~\cite{Brems_2021}. \changes{Skyrmion systems, in particular, allow for superimposing thermal diffusion with artificial diffusion (e.g., current-based~\cite{Brems_2021} or field-based~\cite{Gruber_nodate}) and thus the tokens’ dynamics can be adapted to the circuit geometry such that the tokens find the computational forward paths faster. Moreover, the possibility of significant speed-up at the expense of additional energy can mitigate the disadvantage non-deterministic computation times may pose for time-critical applications.~\cite{Brems_2021}}  Experimental advances have been made by Jibiki et al. in implementing skyrmion-based circuit modules for Brownian token-based computing~\cite{Jibiki_2020,Nozaki_2019}.\\

\section{Perspective of skyrmions for \changes{reservoir} computing}

Skyrmion-based non-conventional computing is an emerging field that aims to harness the distinct properties of magnetic skyrmions. While this field has already shown several promising results and is expected to play a significant role in the future of unconventional computing, there is still a multitude of challenges left to overcome. These include both short-term goals to make skyrmion-based reservoir computing more competitive, practical, and efficient, as well as long-term challenges that need to be addressed to realise the full potential of future skyrmion-based computing.

The field of RC currently features a wide range of designs and architectures with virtually endless potential for further customisation and experimentation since it encompasses any dynamical phenomenon that can be harnessed to build reservoirs. While this diversity is a strength of RC, it also presents one of its biggest challenges as it makes it difficult to achieve cohesion and standardisation among the different systems. This is also evident in skyrmion RC where a wide range of diverse system designs have already been put forward, as elaborated in Section 3, despite the field being in its nascent stages. For example, there are practically no universally applied measures or standards for evaluating and comparing PRC models. Such methods would not only facilitate the selection of the most suitable models for a specific task but also grant researchers valuable insight into the underlying principles and dynamics of reservoir systems. This understanding can be further leveraged to create unconventional systems that incorporate the most advantageous features of existing models. Additionally, by comparing models with varying parameters such as reservoir size and connectivity, researchers can determine the effect of these parameters on system performance and optimise the design of the reservoir accordingly. Although we are yet to realise a unified formalism for RC, progress is being made in certain sub-fields of PRC to introduce reliable measures.
In the particular case of skyrmion RC, task-agnostic local metrics have been proposed~\cite{Love_2021}. In addition to performance classifications, it is also important to establish what constitutes fair comparison among models. There have been instances in PRC where researchers have selected or modified datasets to achieve favourable benchmark results. Implementing standardised comparison schemes would help to eliminate such practices and ensure fair and unbiased evaluations. 

Reservoirs possess memory capabilities that enable them to retain information about past inputs for a certain period of time due to recurrent connections within the reservoir~\cite{Verstraeten_2007}. However, nonlinearity in the reservoir's dynamics decreases memory capacity~\cite{Inubushi_2017}. Therefore, the role of nonlinearity, which is necessary for input mapping, needs to be balanced with the memory capacity for optimal performance. One way to achieve this in RC design is to create a gradient mixture reservoir, where one section of the reservoir has high nonlinearity/low memory, while another section has high memory/low nonlinearity~\cite{Stenning_2022}. Additionally, it is possible to achieve output states that emphasise either memory or nonlinearity by strategically placing readout contacts~\cite{Love_2021}.

In task-specific RC, it will be important to identify tasks that are particularly suited to skyrmion-based reservoirs. These tasks will fundamentally depend on the intrinsic properties of skyrmion systems, such as the time scales of driven dynamics and decorrelation dynamics (fading memory) as well as the nonlinearity of skyrmion interactions. 
One of the reasons skyrmions have emerged as promising physical reservoir candidates is their internal and collective dynamics on different time scales and competing interactions on different length scales. \changes{Timescale harmonisation plays a crucial role in effectively harnessing these distinctive properties, ensuring that diverse temporal behaviours are synchronised and optimally integrated.} Time scales are particularly important to consider since a computationally expensive conversion algorithm from the relevant time scale of a specific task to that of the reservoir may prove to be a bottleneck and potentially impede the high computational speed and energy efficiency advantages of RC.

Another important factor to take into consideration is the plethora of excitation methods for skyrmion dynamics like spin-torques~\cite{Jonietz_2010,Yu_2012,Woo_2016, Litzius_2020, Litzius_2017} and field gradients~\cite{Moutafis_2009,Buttner_2015} and readout methods like magneto-resistance~\cite{Tomasello_2017}. These methods must be gauged with regard to their applicability in skyrmion-based RC. Apart from stimulation and response measurement methods, the skyrmion pinning effect has played a significant role in recent RC concepts. Skyrmion pinning is an essential ingredient for some RC approaches and an obstacle for others. So methods to engineer the strength and distribution of pinning areas and thus tune it as necessary for a given reservoir may be major benefits for future RC approaches. This includes both material engineering as well as methods to manipulate the effective pinning effect on state-of-the-art samples~\cite{Gruber_nodate}. Finally, the trade-offs introduced by thermal effects in skyrmion reservoirs must be further investigated. It has been demonstrated that thermal dynamics can benefit a device’s energy efficiency and error tolerance. On the downside, the stochasticity accompanying thermal dynamics is expected to act to the detriment of the systems’ short- and long-term memory.

\begin{figure*}[t]
\includegraphics[width=\linewidth]{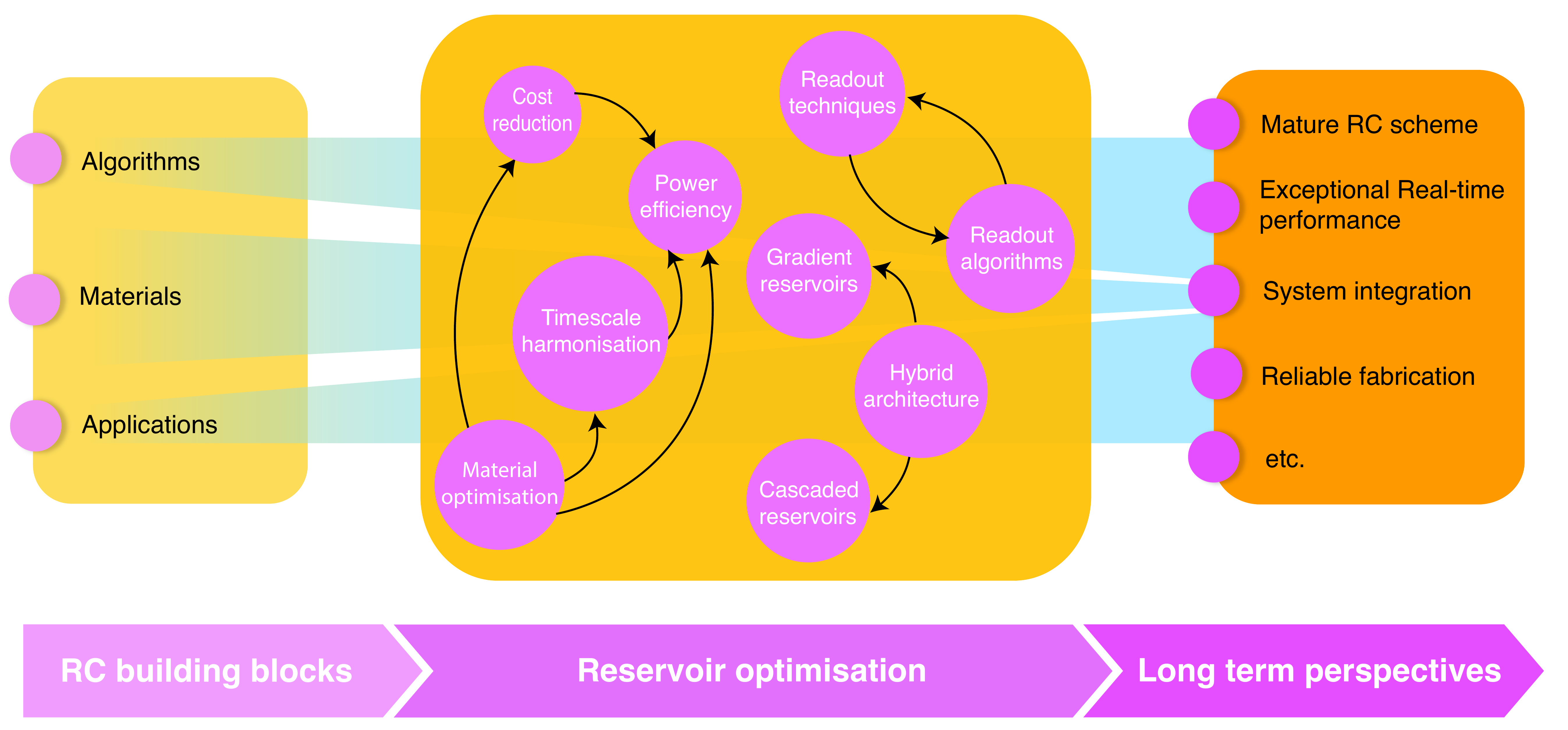}

    \caption{Skyrmion reservoir computing perspectives.}
    \label{fig:perspectives}

\end{figure*}

The future of skyrmion RC holds significant promise, with numerous untapped research avenues to be explored, including skyrmion oscillators, cyclic reservoirs, and beyond. Figure~\ref{fig:perspectives} depicts the predicted importance of materials, algorithms, and applications, visualised by blue-coloured beams that vary in size over time. As research progresses, the field will likely move from a broad range of reservoir materials to only a narrow selection of the most effective, while algorithms and applications are anticipated to grow with advancements in research. 
\changes{In particular, architectures will presumably evolve beyond standard RC. For example, the incorporation of the Brownian computing paradigm was an initiative to blend different ideas into existing RC concepts.}
Additionally, the figure \changes{suggests some key areas for reservoir optimisation, such as the use of cascaded architectures, in which multiple physical reservoirs are connected sequentially, allowing for the enhancement of computational capacity and the representation of more complex tasks\cite{Stenning_2022}. This approach optimises the spatial and temporal characteristics of the input data by leveraging the advantages of each individual reservoir. One can also combine the unique strengths of RC with other computational approaches, offering a powerful and versatile hybrid framework for solving complex tasks. For instance, RC can be combined with ANNs to enhance the capacity to learn intricate patterns and generalise effectively~\cite{Tong_2018}. By fusing RC's memory-enhanced capabilities and rapid adaptation with the robust learning and optimisation mechanisms of ANNs, an ANN/RC hybrid system can be optimised to tackle a wide range of applications, from time series prediction and signal processing to natural language processing and image recognition. } 

\changes{Scalability is a key property where skyrmion RCs likely excel, surpassing alternatives of individually connected components like MTJs\cite{Finocchio_2021}, nano-oscillators\cite{Romera_2018}, 
artificial nanomagnets\cite{Stenning_2022}, and so forth.
The intrinsic advantage of skyrmion RC over these systems lies in their natural connectivity and a vast number of degrees of freedom, which would otherwise require artificial enhancement. This naturally leads to more efficient spatial packing and fewer energy losses that would result from the overhead caused by the wiring of the individual components. Additionally, skyrmions provide a localised and topologically stabilised alternative to general spin-wave or domain wall-based devices.
Although skyrmions exhibit favourable scalability properties, managing and controlling them effectively becomes increasingly challenging as systems grow larger. To tackle this issue, it is crucial to develop efficient, scalable algorithms tailored for large-scale reservoirs, in addition to other stabilisation mechanisms. These algorithms should be able to adapt to variations in the system's size, complexity, or environment, including readout techniques. Incorporating self-organisation principles may be beneficial, as they allow the system to dynamically reconfigure and maintain its computational capabilities while scaling.}

\changes{Despite having numerous advantages that strengthen their position in the emerging field of neuromorphic computing, skyrmions face further challenges that are yet to be addressed. For instance, the development of efficient skyrmion-based devices requires overcoming difficulties in material engineering and optimisation, as well as the need for more advanced fabrication techniques.} While significant progress has been made in recent years, further research is necessary to identify more affordable and durable material systems that host stable skyrmions. Moreover, it will be crucial to develop both efficient and cost-effective readout methods that can merge seamlessly with I/O components of electronic systems. \changes{Currently, reliable optical readout techniques for skyrmions face considerable obstacles, such as limited spatial resolution, signal-to-noise ratios affected by thermal factors, and constraints on high-speed detection. To circumvent these limitations, one can utilise electrical readout techniques that take advantage of magnetoresistive effects. Such methods are presently employed in skyrmion neuromorphic computing prototypes. However, this approach necessitates a relatively large voltage within the device, calling for overall device size expansion. Moving forward, the development of ultrasensitive detection methods will be crucial.}

\changes{To conclude, skyrmion-based unconventional computing shows great potential as a research area. It is exciting to anticipate the progress of this field over the next few years and to see how skyrmions will be incorporated into mainstream computing applications.}



\vskip 20pt
\bibliography{Bibliography}

\section*{Acknowledgments}
O.L. and H.K. thank the Leverhulme Trust for financial support via RPG-2016-391. M.K., K.E-S, R.M. and M.B. are grateful to the Deutsche Forschungsgemeinschaft (DFG, German Research Foundation) for funding this research: Project numbers \#403502522 and \#403233384-SPP 2137 Skyrmionics and \#320163632 (Emmy Noether) and acknowledge funding from the Emergent AI Centre funded by the Carl-Zeiss-Stiftung.
M.K., K.E.S and M.B. also acknowledge funding from TopDyn and SFB TRR 173 Spin+X (project A01 \#268565370 and project B12 \#268565370. M.K., and M.B. further acknowledge funding from the Horizon 2020 Framework Programme of the European Commission under FET-Open Grant No. 863155 (s-Nebula), Grant No. 856538 (ERC-SyG 3D MAGIC), No. 101070290 (\changes{HORIZON-CL4-2021-DIGITAL-EMERGING-01-14 }NIMFEIA) and under the Marie Skłodowska-Curie Grant Agreement No. 860060 “Magnetism and the effect of Electric Field” (MagnEFi). M.B. is supported by a doctoral scholarship of the Studienstiftung des deutschen Volkes and acknowledges funding from TRR 146 (project \#233630050).



\end{document}